\documentclass[10pt,aps,pra,twocolumn,superscriptaddress]{revtex4-2}
\usepackage{amsmath}
\usepackage{amssymb}
\usepackage{graphicx}
\usepackage{hyperref}
\usepackage[english]{babel}
\usepackage{babel}
\usepackage[usenames,dvipsnames]{color}

\begin{document}

\title{Generation of photons from vacuum in cavity via time-modulation of a
qubit invisible to the field}
\author{M. V. S. de Paula}
\affiliation{Instituto de F\'{\i}sica, Universidade de Bras\'{\i}lia, Caixa Postal 04455, CEP
70919-970 Bras\'{\i}lia, DF, Brasil}
\author{W. W. T. Sinesio}
\affiliation{Instituto de F\'{\i}sica, Universidade de Bras\'{\i}lia, Caixa Postal 04455, CEP
70919-970 Bras\'{\i}lia, DF, Brasil}
\author{A. V. Dodonov}
\affiliation{Centro Internacional de F\'{\i}sica (CIF), Instituto de F\'{\i}sica, Universidade de Bras\'{\i}lia, Caixa Postal 04455, CEP
70919-970 Bras\'{\i}lia, DF, Brasil}

\begin{abstract}
We propose a scheme for generation of photons from vacuum due to
time-modulation of a quantum system coupled indirectly to the cavity field
through some ancilla quantum subsystem. We consider the simplest case when
the modulation is applied to an artificial 2-level atom (we call t-qubit),
while the ancilla is a stationary qubit coupled via the dipole interaction
both to the cavity and t-qubit. We find that tripartite entangled states
with a small number of photons can be generated from the system ground state
under resonant modulations, even when the t-qubit is far detuned from both
the ancilla and the cavity, provided its bare and modulation frequencies are
properly adjusted as function of other system parameters. We attest our
approximate analytic results by numeric simulations and show that photon
generation from vacuum persists in the presence of common dissipation
mechanisms.
\end{abstract}

\maketitle

\section{Introduction}

Dynamical Casimir effect (DCE) designates a plethora of phenomena
characterized by generation of photons (or quanta of some other field) from
vacuum due to time-dependent variations of the geometry (dimensions) or
material properties (e. g., the dielectric constant or conductivity) of some
macroscopic system (see, e. g., the reviews \cite{rev1,rev2,rev3,rev4,rev5}%
). It was initially investigated for Electromagnetic (EM) field in the
presence of nonuniformly accelerating mirrors and cavities with oscillating
boundaries or time-dependent material properties \cite%
{ful,maia,moore,law,lambe}, but the concept was later extended to
optomechanical systems \cite{opto1,opto2}, Bose-Einstein condensates and
ultracold gases \cite{bec1,bec2,bec3}, polariton condensates \cite{12} and
spinor condensates \cite{11,11a}. Recently, DCE was implemented
experimentally via periodical fast changes of the boundary conditions in
circuit Quantum Electrodynamics architecture (circuit QED) \cite%
{Wilson11-Nature,Johan13,Paraoanu13,Svensson18} and Bose-Einstein
condensates \cite{13}. Beside serving as a direct proof of the vacuum
fluctuations \cite{rev5}, from the practical point of view DCE can be
employed to generate nonclassical states of light or ensemble of atoms \cite%
{nc1,nc2,Johan13}.

The circuit QED architecture \cite{c1,c2,c3,c4,c5} is a handy platform for
the implementation of DCE and its generalizations, since both the cavity's
and artificial atoms' properties can be rapidly modulated by external bias,
e. g., magnetic flux \cite{bead,silveri}. In particular, when the atom is
directly coupled to the field via the dipole interaction (described by the
Quantum Rabi Model \cite{m1}), a resonant time-modulation of the atomic
transition frequency or the atom-field coupling strength can be used to
generate photons and light--matter entangled states from the initial vacuum
state \cite{AVD09,AVD14,simone,ibe}. In this case, one can view the atom as
a microscopic constituent of the intracavity medium that shifts its
effective frequency; moreover, such scheme benefits from leaving the Fock
states of the cavity field time-independent (as opposed to the standard case
of time-varying cavity frequency, when the annihilation operator and the
Fock states depend explicitly on time \cite{law}). These nonstationary
circuit QED setups exhibit several important phenomena beside photon
generation from vacuum, e. g.: generation of atom-field entangled states and
novel nonclassical states of light \cite{disc,Rossatto16}, quantum
simulation \cite{Felicetti15,Corona16,Sabin17superlum}, implementation of
quantum gates \cite{gates}, engineering of effective interactions \cite%
{milit}, implementation of quantum thermal engines \cite{werlang,werlang2},
photon generation and atom-field effective coupling via multi-photon
transitions \cite{AVD18,AVD19}, anti-dynamical Casimir effect (coherent
annihilation of excitations due to external modulation) \cite%
{AVD15souza,AVD15,guevara,adcen}, photon generation by both temporal and
spatial modulation in metamaterials \cite{Ma19}, vacuum Casimir-Rabi
oscillations in optomechanical systems \cite{opt}, etc \cite{rev5}.

In this paper we investigate whether photons can also be generated from
vacuum by modulating an artificial atom that is \textquotedblleft
invisible\textquotedblright\ to the cavity field, i. e., not directly
coupled to the field. Instead, it is coupled to the cavity field indirectly
through some auxiliary subsystem -- \textit{ancilla}. Such coupling scheme
may have several reasons and applications. For instance, the artificial atom
can be designed specifically to withstand fast external modulation of
arbitrary format, at the expense of null coupling to the
cavity field but large coupling to some other subsystems; or the atom can be
placed outside or at the end of the cavity (at the node of the electric
field) to minimize the influence of external driving on the cavity field and
increase the cavity quality factor. In addition, the modulated artificial
atom can be designed to be able to couple selectively to multiple cavities
by means of different stationary ancillas, which are constructed with
reduced dissipative losses and enhanced atom-field coupling strengths
(ultrastrong coupling \cite{m1,m2}, for instance). Independently of the
concrete scenario, it seems timely and important to investigate whether such
\textquotedblleft invisible\textquotedblright\ time-modulated atom can be
harnessed to generate photons from vacuum or engineer some useful
effective interactions, and under which conditions these processes are
optimized. We address analytically and numerically this issue by
considering the simplest scenario in which the time-modulated artificial
atom is a qubit (\textquotedblleft \textit{t-qubit}\textquotedblright , for
shortness) and the ancilla is a stationary qubit dipole-coupled to both the
cavity field and the t-qubit. We find that photon generation with
sufficiently large rates is possible provided there is a fine tuning of both
the modulation frequency and the bare frequency of the t-qubit, which depend
on all other system parameters.

This paper is organized as follows. The mathematical formulation of our
proposal is given in Section \ref{fom}, and in Section \ref{ana} we present
a closed analytic description of the dynamics in terms of the system
dressed-states. In Section \ref{nume} we confirm our analytic predictions by
exact numeric simulations and illustrate typical system behavior in
different regimes of operation. In particular, we show that the initial
vacuum state can be deterministically driven either to states with only two
excitations or states with multiple excitations, even in the presence of
weak dissipative effects. Section \ref{conc} contains the conclusions.

\section{Mathematical formulation\label{fom}}

\begin{figure}[tbh]
\begin{center}
\includegraphics[width=0.49\textwidth]{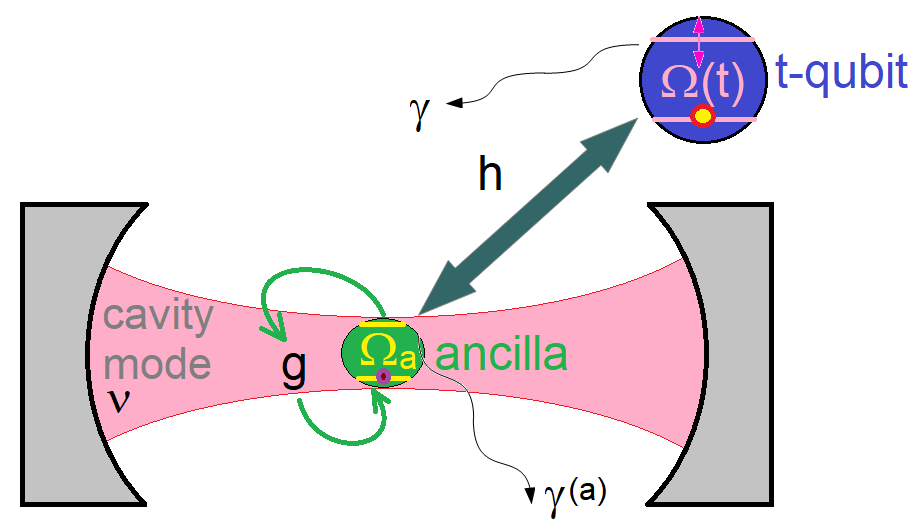} {}
\end{center}
\caption{Schematic of the proposal. Extracavity \textit{t-qubit} [of
time-dependent frequency $\Omega (t)$] is coupled to the \textit{ancilla}
(of frequency $\Omega _{a}$) with coupling strength $h$. The ancilla is
coupled to the cavity mode (of frequency $\protect\nu $) with coupling
constant $g$. T-qubit (ancilla) also has damping and pure dephasing rates $%
\protect\gamma $ and $\protect\gamma _{ph}$ ($\protect\gamma ^{(a)}$ and $%
\protect\gamma _{ph}^{(a)}$).}
\label{fig0}
\end{figure}

Our tripartite system is represented schematically in Figure \ref{fig0}, and
is described by the Hamiltonian ($\hbar =1$)%
\begin{equation}
\hat{H}=\left[ \nu \hat{n}+\Omega (t)\hat{\sigma}_{e}+h\hat{\sigma}_{x}\hat{%
\sigma}_{x}^{\left( a\right) }\right] +\left[ \Omega _{a}\hat{\sigma}%
_{e}^{\left( a\right) }+g\left( \hat{a}+\hat{a}^{\dagger }\right) \hat{\sigma%
}_{x}^{\left( a\right) }\right] \,.  \label{tot}
\end{equation}%
The terms in the first brackets describe the free cavity field and the
t-qubit coupled to the ancilla, while the terms in the second brackets
describe the ancilla qubit and its coupling to the cavity field. $\nu $ is
the cavity frequency, $\hat{n}=\hat{a}^{\dagger }\hat{a}$ is the photon
number operator and $\hat{a}$ ($\hat{a}^{\dagger }$) is the annihilation
(creation) operator. The t-qubit operators are $\hat{\sigma}_{e}=|e\rangle
\langle e|$, $\hat{\sigma}_{z}=|e\rangle \langle e|-|g\rangle \langle g|$, $%
\hat{\sigma}_{-}=|g\rangle \langle e|$, $\hat{\sigma}_{+}=\hat{\sigma}%
_{-}^{\dagger }$ and $\hat{\sigma}_{x}=\hat{\sigma}_{+}+\hat{\sigma}_{-}$,
where $|g\rangle $ ($|e\rangle $) denotes the ground (excited) state. For
the ancilla the operators are similar and are indicated by the upper index (a),
while its ground and excited states are denoted as $|g_{a}\rangle $ and $%
|e_{a}\rangle $, respectively. We assume that the ancilla, with constant
transition frequency $\Omega _{a}$, interacts directly with the cavity field
via the dipole interaction with the time-independent coupling strength $g$.
The t-qubit is not directly coupled to the cavity field; instead, it
interacts with the ancilla via the dipole interaction with the coupling
constant $h$. Here we derive closed (approximate) analytic description for
moderate coupling strengths, $g,h\lesssim 0.05\nu $, but our numeric
simulations also explore some interesting phenomena in the ultrastrong
coupling regime with $g,h\sim 0.1-0.3\nu $. Notice that the direct
qubit-qubit coupling occurs naturally in many circuit QED setups \cite%
{dir1,dir2,dir3,dir4}. A related case in which the t-qubit is coupled
simultaneously to two cavity modes was recently studied in \cite{last}.

We assume that the transition frequency of the t-qubit is modulated
externally as%
\begin{equation*}
\Omega (t)=\Omega _{0}+\varepsilon \sin \left( \eta t\right) \,,
\end{equation*}%
where $\Omega _{0}$ is the bare (average) frequency, $\varepsilon \ll
\Omega _{0}$ is the modulation amplitude and $\eta $ is the frequency of
modulation. The generalization of our scheme to nonharmonic modulations is
straightforward by using the Fourier decomposition \cite{AVD14}. We also
notice that time-dependent modulation frequency $\eta (t)$ can originate
novel dynamical behaviors and optimize the photon generation process \cite%
{AVD09,AVD16}, but these regimes lie outside the scope of this paper.

In circuit QED the Hamiltonian alone does not describe accurately the system
dynamics due to the coupling to the environment, so instead of the Schr\"{o}%
dinger equation (SE) one has to use the master equation for the density operator $%
\hat{\rho}$%
\begin{equation}
\frac{d{\hat{\rho}}}{dt}=-i[\hat{H},\hat{\rho}]+\mathcal{\hat{L}}\hat{\rho},
\label{ssme1}
\end{equation}%
where the Liouvillian $\mathcal{\hat{L}}$ accounts for the influence of the
environment \cite{vogel,schleich}. The form of the Liouvillian depends on
the spectral density of the reservoir and the type of coupling to the system
\cite{Breuer}. For moderate coupling strengths one can use the standard
Markovian master equation (SMME) of quantum optics \cite{Carm}%
\begin{equation}
\mathcal{\hat{L}}=\mathcal{\hat{L}}_{d}+\mathcal{\hat{L}}_{ph}+\mathcal{\hat{%
L}}_{d}^{(a)}+\mathcal{\hat{L}}_{ph}^{(a)},  \label{kernel}
\end{equation}%
where the superoperator $\mathcal{\hat{L}}_{d}$ ($\mathcal{\hat{L}}_{ph}$)
describes the energy damping (pure dephasing) of the t-qubit by a thermal
reservoir; $\mathcal{\hat{L}}_{d}^{(a)}$ and $\mathcal{\hat{L}}_{ph}^{(a)}$
have similar meaning for the ancilla. Indeed, it was shown in \cite%
{AVD15,AVD16} that in similar setups the discrepancy between this equation
and a more rigorous one (a microscopic derivation taking into account the
qubit--resonator coupling \cite{gambetta}) is small, while the qualitative
agreement is excellent.

In this paper we consider the zero-temperature limit of the SMME%
\begin{equation}
\mathcal{\hat{L}}_{d}=\gamma \mathcal{D}[\hat{\sigma}_{-}],~\,\mathcal{\hat{L%
}}_{ph}=\frac{\gamma _{ph}}{2}\mathcal{D}[\hat{\sigma}_{z}]\,,  \label{kik2}
\end{equation}%
where $\gamma $ ($\gamma _{ph}$) is the atom relaxation (pure dephasing)
rate and $\mathcal{D}[\hat{\Phi}]\hat{\rho}\equiv (2\hat{\Phi}\hat{\rho}\hat{%
\Phi}^{\dagger }-\hat{\Phi}^{\dagger }\hat{\Phi}\hat{\rho}-\hat{\rho}\hat{%
\Phi}^{\dagger }\hat{\Phi})/2$ is the so called Lindblad superoperator, that
preserves the hermiticity, normalization and positivity of $\hat{\rho}$ \cite%
{Breuer}. For numeric simulations in Section \ref{nume} we shall adopt the
following parameters: $\gamma =5\times 10^{-3}g$, $\gamma _{ph}=\gamma /2$, $%
\gamma ^{(a)}=\gamma /5$ and $\gamma _{ph}^{(a)}=\gamma _{ph}/5$. This
represents the presumable situation in which the t-qubit is exposed to
moderate dissipation due to external modulation, while the ancilla is less
susceptible to dissipation by proper design; these dissipative rates were
assumed sufficiently small, yet readily achievable experimentally \cite%
{63,64,65}. Such approximate treatment is sufficient to assess the
feasibility of our scheme in realistic situations. We also neglected the
cavity damping, assuming that for stationary cavity the losses are
negligible during the time interval of interest.

\section{Analytic description \label{ana}}

The analysis is simplified by introducing a new \textquotedblleft
conjoint\textquotedblright\ atomic basis $\{|A_{i}\rangle ,i=0,...,3\}$, in
which $|A_{i}\rangle $ are the eigenstates of the two-atom time-independent
Hamiltonian $\hat{H}_{a}=\Omega _{0}\hat{\sigma}_{e}+\Omega _{a}\hat{\sigma}%
_{e}^{\left( a\right) }+h\hat{\sigma}_{x}\hat{\sigma}_{x}^{\left( a\right) }$%
containing the bare t-qubit frequency $\Omega _{0}$:
\begin{eqnarray}
|A_{0}\rangle  &=&N_{0}\left[ \left( W_{+}+D_{+}\right) |g,g_{a}\rangle
-h|e,e_{a}\rangle \right]   \label{no} \\
|A_{1}\rangle  &=&N_{1}\left[ \left( W_{-}-D_{-}\right) |g,e_{a}\rangle
+h|e,g_{a}\rangle \right]   \notag \\
|A_{2}\rangle  &=&N_{2}\left[ \left( W_{-}+D_{-}\right) |g,e_{a}\rangle
+h|e,g_{a}\rangle \right]   \notag \\
|A_{3}\rangle  &=&N_{3}\left[ \left( W_{+}-D_{+}\right) |g,g_{a}\rangle
-h|e,e_{a}\rangle \right] \,.  \notag
\end{eqnarray}%
Here $W_{\pm }=(\Omega _{a}\pm \Omega _{0})/2$, $D_{\pm }=\sqrt{W_{\pm
}^{2}+h^{2}}$ and $N_{i}$ are the normalization constants (with the
dimension 1/frequency). The corresponding eigenergies are%
\begin{eqnarray}
\lambda _{0} &=&W_{+}-D_{+}~,~\lambda _{1}=W_{+}-D_{-} \\
\lambda _{2} &=&W_{+}+D_{-}~,~\lambda _{3}=W_{+}+D_{+}\,.  \notag
\end{eqnarray}%
In particular, for $W_{+}\gg h$ the non-normalized states read approximately
$|A_{0}\rangle \approx |g,g_{a}\rangle -(h/2W_{+})|e,e_{a}\rangle $ and $%
|A_{3}\rangle \approx -|e,e_{a}\rangle -(h/2W_{+})|g,g_{a}\rangle $. For $%
\left\vert W_{-}\right\vert \ll h$ we have $|A_{1}\rangle \approx
|e,g_{a}\rangle -|g,e_{a}\rangle $ and $|A_{2}\rangle \approx
|e,g_{a}\rangle +|g,e_{a}\rangle $, while for $\left\vert W_{-}\right\vert
\gg h$%
\begin{equation}
|A_{1}\rangle \approx \left\{
\begin{array}{c}
|e,g_{a}\rangle -(h/2|W_{-}|)|g,e_{a}\rangle ~,~\Omega _{a}>\Omega _{0} \\
-|g,e_{a}\rangle +(h/2|W_{-}|)|e,g_{a}\rangle ~,~\Omega _{a}<\Omega _{0}%
\end{array}%
\right.
\end{equation}%
\begin{equation}
|A_{2}\rangle \approx \left\{
\begin{array}{c}
|g,e_{a}\rangle +(h/2|W_{-}|)|e,g_{a}\rangle ~,~\Omega _{a}>\Omega _{0} \\
|e,g_{a}\rangle +(h/2|W_{-}|)|g,e_{a}\rangle ~,~\Omega _{a}<\Omega _{0}\, .%
\end{array}%
\right.
\end{equation}

In the basis $\{|A_{i}\rangle \}$ the time-independent system Hamiltonian
reads%
\begin{equation}
\hat{H}_{0}=\nu \hat{n}+\sum_{i=0}^{3}\lambda _{i}|A_{i}\rangle \langle
A_{i}|+g\left( \hat{a}+\hat{a}^{\dagger }\right) \hat{\sigma}_{x}^{\left(
a\right) }\,
\end{equation}%
and the total Hamiltonian is $\hat{H}=\hat{H}_{0}+\varepsilon \sin \left(
\eta t\right) \hat{\sigma}_{e}$.

\subsection{Spectrum far from degeneracies}\label{lm1}

Far from any degeneracy between the eigenenergies of $\hat{H}_{0}(g=0)$ (system
Hamiltonian without the matter-field coupling) the spectrum of $\hat{H}_{0}$
can be found from the nondegenerate perturbation theory. We use the
orthonormal complete basis $|A_{i}^{n}\rangle \equiv |A_{i}\rangle \otimes
|n\rangle $ (where $|n\rangle $ is the Fock state of the field with $n\geq 0$
and $i=0,\ldots ,3$) and denote the eigenstate (eigenvalue) of $\hat{H}_{0}$
with the \emph{dominant contribution} of the state $|A_{i}^{n}\rangle $ as $%
|A_{i},n\rangle $ ($\lambda _{i,n}$). For example, to the second order in $g$
one obtains the (non-normalized) state%
\begin{eqnarray}
&&|A_{0},n\rangle =|A_{0}^{n}\rangle \notag \\
&& +g\left[ \frac{\sigma _{01}\sqrt{n}}{%
v-D_{+}+D_{-}}|A_{1}^{n-1}\rangle -\frac{\sigma _{01}\sqrt{n+1}}{%
v+D_{+}-D_{-}}|A_{1}^{n+1}\rangle \right.   \notag \\
&&\left. +\frac{\sigma _{02}\sqrt{n}}{v-D_{+}-D_{-}}|A_{2}^{n-1}\rangle -%
\frac{\sigma _{02}\sqrt{n+1}}{v+D_{+}+D_{-}}|A_{2}^{n+1}\rangle \right]
\label{pt} \\
&&+g^{2}\left[ \sqrt{n(n-1)}\left( \Xi _{n}^{\left( 1\right)
}|A_{0}^{n-2}\rangle +\Xi _{n}^{\left( 4\right) }|A_{3}^{n-2}\rangle \right)
\right.   \notag \\
&&\left. +\sqrt{(n+1)(n+2)}\left( \Xi _{n}^{\left( 2\right)
}|A_{0}^{n+2}\rangle +\Xi _{n}^{\left( 5\right) }|A_{3}^{n+2}\rangle \right)
+\Xi _{n}^{\left( 3\right) }|A_{3}^{n}\rangle \right] \,,  \notag
\end{eqnarray}%
where $\sigma _{ij}\equiv \langle A_{i}|\hat{\sigma}_{x}^{(a)}|A_{j}\rangle $%
, $\sigma _{01}^{2}+\sigma _{02}^{2}=1$ and%
\begin{eqnarray*}
\Xi _{n}^{\left( 1\right) } &=&\frac{1}{2v}\left( \frac{\sigma _{01}^{2}}{%
v-D_{+}+D_{-}}+\frac{\sigma _{02}^{2}}{v-D_{+}-D_{-}}\right)  \\
\Xi _{n}^{\left( 2\right) } &=&\frac{1}{2v}\left( \frac{\sigma _{01}^{2}}{%
v+D_{+}-D_{-}}+\frac{\sigma _{02}^{2}}{v+D_{+}+D_{-}}\right)  \\
\Xi _{n}^{\left( 3\right) } &=&-\frac{\sigma _{01}\sigma _{02}D_{-}}{D_{+}}%
\left( \frac{n}{\left( v-D_{+}+D_{-}\right) \left( v-D_{+}-D_{-}\right) }\right. \\
&&\left. +%
\frac{n+1}{\left( v+D_{+}+D_{-}\right) \left( v+D_{+}-D_{-}\right) }\right)
\\
\Xi _{n}^{\left( 4\right) } &=&\frac{\sigma _{01}\sigma _{02}D_{2}^{-}}{%
\left( v-D_{+}\right) \left( v-D_{+}-D_{-}\right) \left(
v-D_{+}+D_{-}\right) } \\
\Xi _{n}^{\left( 5\right) } &=&-\frac{\sigma _{01}\sigma _{02}D_{2}^{-}}{%
\left( v+D_{+}\right) \left( v+D_{+}+D_{-}\right) \left(
v+D_{+}-D_{-}\right) }\,.
\end{eqnarray*}%
The corresponding eigenenergy reads%
\begin{eqnarray}
\lambda _{0,n} &=&\lambda _{0}+vn+g^{2}\left[ \frac{\sigma _{01}^{2}n}{%
v-D_{+}+D_{-}}-\frac{\sigma _{01}^{2}\left( n+1\right) }{v+D_{+}-D_{-}}%
\right.   \notag \\
&&+\left. \frac{\sigma _{02}^{2}n}{v-D_{+}-D_{-}}-\frac{\sigma
_{02}^{2}\left( n+1\right) }{v+D_{+}+D_{-}}\right] \,.
\end{eqnarray}%
Expressions for other states $|A_{i},n\rangle $ and eigenenergies $\lambda
_{i,n}$ can be obtained similarly, but they will not be necessary for the
present work.

\subsection{Spectrum near degeneracies}

\label{fop}

The easiest applications of our scheme explore the regime of parameters for
which two states of the subspace $\mathcal{A}_{n}=\{|A_{0}^{n}\rangle $,$%
|A_{1}^{n-1}\rangle $, $|A_{2}^{n-1}\rangle $,$|A_{3}^{n-2}\rangle \}$ are
nearly degenerate, as occurs for $D_{+}\pm D_{-}\approx \nu $ (meaning $%
\Omega _{0}\approx \nu $ or $\Omega _{a}\approx \nu $) or $\Omega
_{a}+\Omega _{0}\approx 2\sqrt{\nu ^{2}-h^{2}}$. In this case the
nondegenerate perturbation theory fails, but one can obtain excellent
analytic results by expanding $\hat{H}_{0}$ in the subspace $\mathcal{A}_{n}$
as%
\begin{equation}
\Upsilon _{4\times 4}=X\,\hat{I}+M_{1}~,  \label{matrix}
\end{equation}%
where $\hat{I}$ is the 4x4 identity matrix,%
\begin{equation}
M_{1}=\left(
\begin{array}{cccc}
0 & a & b & 0 \\
a & x & 0 & -c \\
b & 0 & y & d \\
0 & -c & d & z%
\end{array}%
\right)
\end{equation}%
and $X=\nu n+\lambda _{0}$, $a=g\sqrt{n}\sigma _{01}$, $b=g\sqrt{n}\sigma
_{02}$, $c=g\sqrt{n-1}\sigma _{02}$, $d=g\sqrt{n-1}\sigma _{01}$, $%
x=D_{+}-D_{-}-\nu $, $y=D_{+}+D_{-}-\nu $ and $z=2(D_{+}-\nu )$. The
eigenvalues $\Lambda _{n,i}$ of $M_{1}$ (where the index $n$ specifies the
subspace and $i=1,\ldots ,4$ labels the eigenvalue) are the roots of the
quartic equation%
\begin{equation}
\Lambda _{n,i}^{4}+B\Lambda _{n,i}^{3}+C\Lambda _{n,i}^{2}+D\Lambda
_{n,i}+E=0  \label{vf}
\end{equation}%
with constant coefficients%
\begin{eqnarray}
B &=&-\left( x+y+z\right)   \notag \\
C &=&xy+\left( x+y\right) z-a^{2}-b^{2}-c^{2}-d^{2}  \notag \\
D &=&\left( a^{2}+c^{2}\right) y+\left( b^{2}+d^{2}\right) x+\left(
a^{2}+\allowbreak b^{2}\right) z-xyz  \notag \\
E &=&2abcd+a^{2}d^{2}+b^{2}c^{2}-a^{2}\allowbreak yz-b^{2}xz\,.
\end{eqnarray}%
From the Ferrari's method we obtain%
\begin{eqnarray}
\Lambda _{n,1} &=&-\frac{B}{4}-S-\frac{1}{2}\sqrt{-4S^{2}-2p+\frac{q}{S}}
\notag \\
\Lambda _{n,2} &=&-\frac{B}{4}-S+\frac{1}{2}\sqrt{-4S^{2}-2p+\frac{q}{S}}
\notag \\
\Lambda _{n,3} &=&-\frac{B}{4}+S-\frac{1}{2}\sqrt{-4S^{2}-2p-\frac{q}{S}}
\notag \\
\Lambda _{n,4} &=&-\frac{B}{4}+S+\frac{1}{2}\sqrt{-4S^{2}-2p-\frac{q}{S}}\,,
\end{eqnarray}%
where
\begin{equation*}
S=\sqrt{\frac{\Delta _{0}\cos \varphi -p}{6}}\quad ,\quad q=D+\frac{B}{2}%
\left( \frac{B^{2}}{4}-C\right)
\end{equation*}%
\begin{equation}
\varphi =\frac{1}{3}\arccos \left( \frac{\Delta _{1}}{2\Delta _{0}^{3}}%
\right) ~,~p=C-\frac{3}{8}B^{2}
\end{equation}%
\begin{equation*}
\Delta _{0}=\sqrt{C^{2}-3BD+12E}
\end{equation*}
\begin{equation*}
\Delta _{1}=2C^{3}-9C\left(
BD+8E\right) +27\left( B^{2}E+D^{2}\right) \,.
\end{equation*}

The normalized eigenstate corresponding to the eigenvalue $\Lambda _{n,i}$ is%
\begin{equation}
|\phi _{n,i}\rangle =\phi _{n,i}^{\left( 0\right) }|A_{0}^{n}\rangle +\phi
_{n,i}^{\left( 1\right) }|A_{1}^{n-1}\rangle +\phi _{n,i}^{\left( 2\right)
}|A_{2}^{n-1}\rangle +\phi _{n,i}^{\left( 3\right) }|A_{3}^{n-2}\rangle \,,
\label{weight}
\end{equation}%
where $i=1,\ldots ,4$, $\phi _{n,i}^{\left( k\right) }$ are the probability
amplitudes of the conjoint atomic state $|A_{k}\rangle $ given by%
\begin{equation}
\phi _{n,i}^{\left( 0\right) }=\frac{\Theta _{n,i}}{a}\left\{ c-\frac{%
x-\Lambda _{n,i}}{c}\left[ \left( z-\Lambda _{n,i}\right) +d\Phi _{n,i}%
\right] \right\}
\end{equation}%
\begin{equation}
\phi _{n,i}^{\left( 1\right) }=\frac{\Theta _{n,i}}{c}\left[ d\Phi
_{n,i}+z-\Lambda _{n,i}\right] ~,~\phi _{n,i}^{\left( 2\right) }=\Theta
_{n,i}\Phi _{n,i}~,~\phi _{n,i}^{\left( 3\right) }=\Theta _{n,i}
\end{equation}%
and%
\begin{equation}
\Phi _{n,i}=\frac{\left[ c-\left( x-\Lambda _{n,i}\right) \left( z-\Lambda
_{n,i}\right) /c\right] \Lambda _{n,i}/a-a\left( z-\Lambda _{n,i}\right) /c}{%
b+\left[ \Lambda _{n,i}\left( x-\Lambda _{n,i}\right) d/a+ad\right] /c}
\end{equation}%
\begin{eqnarray}
\Theta _{n,i} &=&\left\{ 1+\Phi _{n,i}^{2}+\frac{\left[ d\Phi
_{n,i}+z-\Lambda _{n,i}\right] ^{2}}{c^{2}}\right.    \\
&&+\left. \frac{\left[ c^{2}-\left( x-\Lambda _{n,i}\right) \left( z-\Lambda
_{n,i}\right) -\left( x-\Lambda _{n,i}\right) d\Phi _{n,i}\right] ^{2}}{%
a^{2}c^{2}}\right\} ^{-1/2}\,.\notag
\end{eqnarray}%
The eigenenergy of the state $|\phi _{n,i}\rangle $ is denoted as $\lambda
_{n,i}^{\phi }$ and reads $\lambda _{n,i}^{\phi }=X+\Lambda _{n,i}$. For
example, according to our notation, near the degeneracy point the eigenstate
$|A_{0},n\rangle $ of $\Upsilon _{4x4}$ is one of the four states $|\phi
_{n,i}\rangle $ for which $\phi _{n,i}^{\left( 0\right) }$ is the largest;
accordingly, the eigenenergy $\lambda _{0,n}$ is one of the four functions
of $\lambda _{n,i}^{\phi }$ (so $\lambda _{0,n}$ can be a discontinuous
function of parameters near the degeneracy point).

The above diagonalization procedure has one major drawback: it neglects the
coupling of the subspace $\mathcal{A}_{n}$ to the neighboring subspaces $%
\mathcal{A}_{n\pm 2}$, carried by the counter-rotating terms $g(\hat{a}\hat{%
\sigma}_{-}^{(a)}+h.c.)$ and $h(\hat{\sigma}_{-}\hat{\sigma}_{-}^{(a)}+h.c.)$
in the Hamiltonian $\hat{H}_{0}$ (although the counter-rotating terms are
fully taken into account within the subspace $\mathcal{A}_{n}$). For small
values of $g$ the main effect of the neglected contributions is to introduce
small frequency shifts (\textquotedblleft Bloch-Siegert\textquotedblright\
shifts \cite{AVD14}) to the eigenfrequencies of the bare Hamiltonian $\hat{H}%
_{0}(g=0)$. To include these corrections in a simplified manner we consider
the subspace $\mathcal{B}_{n}=\{|A_{0}^{n-2}\rangle $,$|A_{1}^{n-1}\rangle $%
, $|A_{2}^{n-1}\rangle $,$|A_{3}^{n}\rangle \}$ containing the basis states
connected solely by the counter-rotating terms. In this basis, the 4x4
matrix corresponding to the Hamiltonian $\hat{H}_{0}$ is $\left[ \nu \left(
n-2\right) +\lambda _{0}\right] \hat{I}+M_{2}$, where%
\begin{equation}
M_{2}=\left(
\begin{array}{cccc}
0 & g\sqrt{n-1}\sigma _{01} & g\sqrt{n-1}\sigma _{02} & 0 \\
g\sqrt{n-1}\sigma _{01} & \nu +D_{+}-D_{-} & 0 & -g\sqrt{n}\sigma _{02} \\
g\sqrt{n-1}\sigma _{02} & 0 & \nu +D_{+}+D_{-} & g\sqrt{n}\sigma _{01} \\
0 & -g\sqrt{n}\sigma _{02} & g\sqrt{n}\sigma _{01} & 2\left( \nu
+D_{+}\right)
\end{array}%
\right)
\end{equation}%
has the same structure as $M_{1}$, so its eigenvalues can be found from the
characteristic equation similar to Equation (\ref{vf}). Denoting the eigenvalues
of $M_{2}$ in the increasing order as $\tilde{\Lambda}_{n,i}$ ($i=1,\ldots ,4
$), the frequency shifts of the states in the subspace $\mathcal{B}_{n}$ are
found as $\delta _{0}^{n-2}\equiv \tilde{\Lambda}_{n,1}$, $\delta
_{1}^{n-1}\equiv \tilde{\Lambda}_{n,2}-\left( \nu +D_{+}-D_{-}\right) $, $%
\delta _{2}^{n-1}\equiv \tilde{\Lambda}_{n,3}-\left( \nu +D_{+}+D_{-}\right)
$ and $\delta _{3}^{n}\equiv \tilde{\Lambda}_{n,4}-2\left( \nu +D_{+}\right)
$ ($\delta _{l}^{k}$ is the frequency shift of the state $|A_{l}^{k}\rangle $%
). Now one can insert these shifts into the matrix (\ref{matrix}) by
replacing $X\rightarrow X+\delta _{0}^{n}$, $x\rightarrow x+\delta
_{1}^{n-1}-\delta _{0}^{n}$, $y\rightarrow y+\delta _{2}^{n-1}-\delta
_{0}^{n}$ and $z\rightarrow z+\delta _{3}^{n-2}-\delta _{0}^{n}$, obtaining
analytically the eigenvalues and eigenstates of the Hamiltonian $\hat{H}_{0}$
near the degeneracy points. As will be shown in Section \ref{nume}, for
moderate coupling strengths $g\lesssim 0.05\nu $ this procedure is in
excellent agreement with exact numeric results.

For higher values of $g$ or direct transitions involving $n>2$ photons,
the above diagonalization procedure becomes improper (since $\mathcal{A}_{n}$
only contains states differing by at most two photons) and must be
generalized by adding more states to the subspace $\mathcal{A}%
_{n}$ (forming a larger subspace $\mathcal{A}_{n}^{gen}$). However, since there is no algebraic
solution to general polynomial equations of degree five or higher with
arbitrary coefficients (Abel--Ruffini theorem), it is easier to perform the
diagonalization numerically, as will be done in Section \ref{fin}.

\subsection{Dynamics in the dressed-basis}

The unitary system dynamics is straightforward in terms of the eigenstates (%
\textit{dressed-states}) of $\hat{H}_{0}$, which can be found either
numerically or analytically as in the previous subsections. Expanding the
wavefunction as%
\begin{equation}
|\psi (t)\rangle =\sum_{l=0}^{\infty }e^{-itE_{l}}A_{l}(t)|\varphi
_{l}\rangle \,,
\end{equation}%
where $|\varphi _{l}\rangle $ and $E_{l}$ are the eigenstates and
eigenvalues of $\hat{H}_{0}$ in the increasing order ($E_{l+1}\geq E_{l}$),
we obtain for the probability amplitudes%
\begin{equation}
i\dot{A}_{m}=\varepsilon \sin \left( \eta t\right) \sum_{l=0}^{\infty
}A_{l}e^{-itE_{lm}}\langle \varphi _{m}|\hat{\sigma}_{e}|\varphi _{l}\rangle
\,,  \label{am}
\end{equation}%
where $E_{lm}=E_{l}-E_{m}$. For the realistic case of weak modulations, $%
\varepsilon \ll \eta $, one can neglect the rapidly oscillating terms to
obtain%
\begin{equation}
\dot{A}_{m}=-\sum_{l\neq m}^{\infty }\text{sign}\left( E_{lm}\right) e^{-%
\text{sign}\left( E_{lm}\right) it\left( \left\vert E_{lm}\right\vert -\eta
\right) }R_{m;l}A_{l}\,,  \label{jk}
\end{equation}%
where
\begin{equation}
R_{m;l}\equiv \frac{\varepsilon }{2}\langle \varphi _{m}|\hat{\sigma}%
_{e}|\varphi _{l}\rangle \,.  \label{rml}
\end{equation}%
Therefore, for the resonant modulation frequency $\eta =\left\vert
E_{lm}\right\vert $ the external perturbation induces transition between the
dressed-states $|\varphi _{m}\rangle $ and $|\varphi _{l}\rangle $ with the
transition rate $R_{m;l}$. From the practical standpoint, the numeric
results are obtained by finding the eigenvalues $E_{m}$ and eigenstates $%
|\varphi _{m}\rangle $ of $\hat{H}_{0}$ in the basis $\{|A_{k}^{n}\rangle \}$%
, where $k=0,...,3$ and $n=0,...,n_{tr}$ ($n_{tr}$ stands for the
maximum number of photons fixed by the truncation procedure, which does
not affect the results for eigenstates with photon numbers $n\ll n_{tr}$),
and then evaluating the transition rates $R_{m;l}$ and energy differences $%
E_{lm}$. For analytic calculations, one simply uses the closed form
expressions for dressed-states and eigenenergies found in Subsections \ref%
{lm1} (for the states $|A_{i},n\rangle $ far from degeneracy points, with
eigenenergies $\lambda _{i,n}$) and \ref{fop} (for dressed-states $|\phi
_{n,i}\rangle $ near degeneracy points, with eigenenergies $\lambda
_{n,i}^{\phi }$) to evaluate the transition rate and energy differences.
Since all the states were expanded in the common basis $\{|A_{k}^{n}\rangle
\}$, such calculations are straightforward.

In this work we are primarily interested in photon generation from the
initial (nondegenerate) ground state of the system $|A_{0},0\rangle $. Near
the degeneracy points one can obtain approximate analytic expression for the
transition rate using the formulae (\ref{pt}) and (\ref{weight}). To the
first order in $g$, the transition rate to the 2-excitations state $|\phi
_{2,i}\rangle $ of subspace $\mathcal{A}_{2}$ is%
\begin{equation}
R_{0,0;2,i}=\frac{\varepsilon h^{2}}{2}\left[ N_{0}N_{3}\phi _{2,i}^{\left(
3\right) }-gT_{i}\right]   \label{rate}
\end{equation}%
\begin{equation}
T_{i}=\left( \frac{\sigma _{01}N_{1}}{v+D_{+}-D_{-}}+\frac{\sigma _{02}N_{2}%
}{v+D_{+}+D_{-}}\right) \left( N_{1}\phi _{2,i}^{\left( 1\right) }+N_{2}\phi
_{2,i}^{\left( 2\right) }\right) \,.
\end{equation}%
Analogously, the transition rate between the dressed-states $|\phi _{n,i}\rangle $
and $|\phi _{n+2,j}\rangle $ is
\begin{equation}
R_{n,i;n+2,j}=\frac{\varepsilon }{2}\langle \phi _{n,i}|\hat{\sigma}%
_{e}|\phi _{n+2,j}\rangle =\frac{\varepsilon }{2}N_{0}N_{3}h^{2}\phi
_{n,i}^{\left( 0\right) }\phi _{n+2,j}^{\left( 3\right) }\,.
\end{equation}%
When the modulation frequency matches only one possible value $\left\vert
E_{lm}\right\vert $ (with nonzero transition rate $R_{m;l}$ between the
respective eigenstates), the dressed-states $|\varphi _{m}\rangle $ and $%
|\varphi _{l}\rangle $ become resonantly coupled and exhibit sinusoidal
behaviors (see Figures \ref{fig1} and \ref{fig2} below): $\left\vert
A_{m}\right\vert ^{2}=\cos ^{2}\left( R_{m;l}t\right) $ and $\left\vert
A_{l}\right\vert ^{2}=\sin ^{2}\left( R_{m;l}t\right) $ (assuming that only $%
A_{m}$ was initially nonzero). On the other hand, when several values $%
\left\vert E_{lm}\right\vert $ are close to the modulation frequency (with
the mismatches $\eta -\left\vert E_{lm}\right\vert $ smaller or of the order
of $R_{m;l}$), then several dressed-states can become simultaneously coupled
and the dynamics becomes more intricate (see Figures \ref{fig3} -- \ref{fig5}%
). We also note that the neglected rapidly oscillating terms introduce small
corrections to the resonant frequencies $|E_{lm}|$ \cite%
{AVD14,AVD15souza}, which are found numerically in this paper.

The ground state can also be directly coupled to dressed-states with more
than two excitations. To obtain reliable analytic formulae for the resonant
modulation frequencies and transition rates one would need to generalize the
results of Section \ref{fop} for larger subspaces (more than four states in $%
\mathcal{A}_{n}^{gen}$). However, we can assure that these transitions do
take place by substituting the formulae (\ref{pt}) and (\ref{weight}) into
Equation (\ref{rml}) to obtain the (overly underestimated) 4-excitations
transition rate%
\begin{eqnarray}
R_{0,0;4,i}&=&\frac{\varepsilon }{2}\langle A_{0},0|\hat{\sigma}_{e}|\varphi
_{4,i}\rangle \\
& \approx& \sqrt{2!}\frac{\varepsilon }{2}h^{2}g^{2}\phi
_{4,i}^{\left( 3\right) }N_{3}\left( N_{0}\Xi _{0}^{\left( 2\right)
}+N_{3}\Xi _{0}^{\left( 5\right) }\right)\notag
\end{eqnarray}%
corresponding to the transition $|A_{0},0\rangle \rightarrow |\phi
_{4,i}\rangle $ of the subspace $\mathcal{A}_{4}$. Similarly, if one
expanded the ground state to the fourth order in $g$ using nondegenerate
perturbation theory in Equation (\ref{pt}), one would obtain $R_{0,0;6,i}\propto
\sqrt{4!}\frac{\varepsilon }{2}h^{2}g^{4}\phi _{6,i}^{\left( 3\right) }$ for
the transition $|A_{0},0\rangle \rightarrow |\phi _{6,i}\rangle $ of the
subspace $\mathcal{A}_{6}$. In Section \ref{fin} we shall calculate the
transition rates for the transition $|A_{0},0\rangle \rightarrow
|A_{0},n\rangle $ with $n=4$ and $6$ by exact numeric diagonalization of the
Hamiltonian $\hat{H}_{0}$; we shall show that these transition rates are
strongly enhanced in the vicinity of degeneracy between the states $%
|A_{0},n\rangle $ and $|A_{2},1\rangle $ or $|A_{3},0\rangle $. Since such
multi-photon transitions are weaker than 2-photons ones, we shall study their implementation in the ultrastrong coupling regime \cite{m1,m2}%
, $g\sim 0.2-0.3\nu $.

Our scheme can be readily generalized to simultaneous modulation of other
system parameters: one merely needs to add the corresponding matrix element
on the RHS of Equation (\ref{am}), which would produce an additive contribution
to the transition rate \ref{rml}. The inclusion of terms proportional to $%
\varepsilon ^{2}$ in Equation (\ref{jk}) is also possible \cite{Silva16}, but is
not considered here because the resulting transition rates are roughly $\eta
/\varepsilon $ times smaller than Equation (\ref{rml}) (although in this case one
benefits from halved resonant modulation frequencies).

\section{Numeric results and discussion \label{nume}}

\begin{figure}[tbh]
\begin{center}
\includegraphics[width=0.49\textwidth]{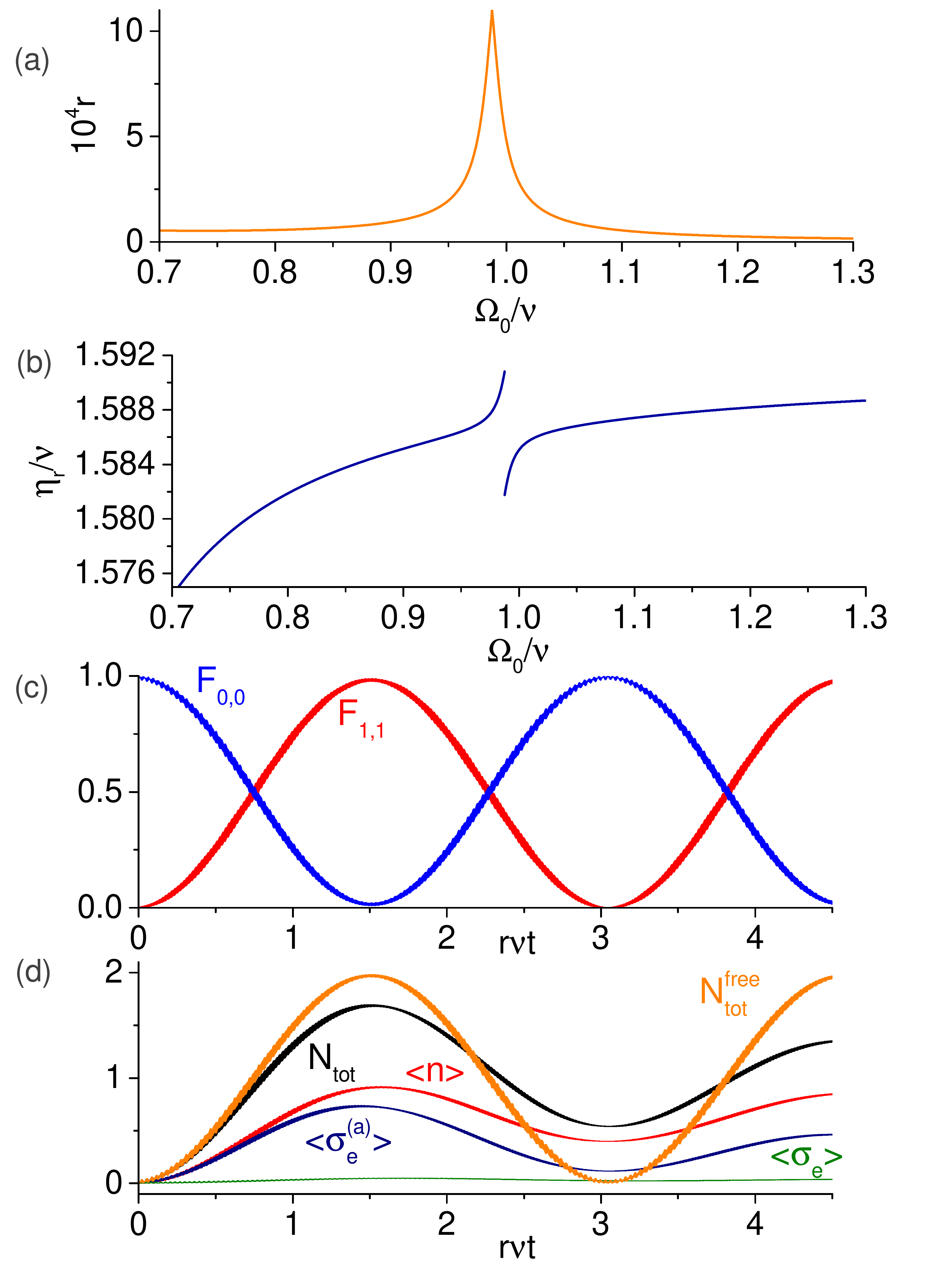} {}
\end{center}
\caption{Numeric results for the transition $|A_{0},0\rangle \leftrightarrow
|A_{1},1\rangle $ using dispersive ancilla with frequency $\Omega _{a}=0.6%
\protect\nu $. a) Dimensionless transition rate $r=|R_{0,0;1,1}|/\protect\nu
$ as function of $\Omega _{0}/\protect\nu $. b) Resonant modulation
frequency, in which the discontinuity arises because the state $%
|A_{1},1\rangle $ is different below and above the degeneracy between the
bare states $|A_{0}^{2}\rangle $ and $|A_{2}^{1}\rangle $. c) Fidelities of
the states $|A_{0},0\rangle $ and $|A_{1},1\rangle $ obtained via numeric
solution of the SE with the total Hamiltonian (\protect
\ref{tot}). As expected, the period of oscillation is $\protect\pi (\protect%
\nu r)^{-1}$. d) Dynamics of the average photon number $\langle n\rangle $,
qubits excitations $\langle \protect\sigma _{e}\rangle $ and $\langle
\protect\sigma _{e}^{(a)}\rangle $, and the total number of excitations $%
N_{tot}$ obtained via numeric solution of the master equation in the
presence of dissipation. For comparison, $N_{tot}^{free}$ illustrates the
average total number of excitations without dissipation, when it attains the
maximum value $N_{tot}^{free}=2$. }
\label{fig1}
\end{figure}

To assess the feasibility of our scheme for photon generation from vacuum in
circuit QED, we first assume conservative experimental parameters $%
g=h=0.05\nu $. In the following, we set the modulation amplitude as $%
\varepsilon =0.1\Omega _{0}$. In Figure \ref{fig1}a we plot the
dimensionless transition rate $r\equiv |R_{0,0;1,1}|/\nu $ for transition
from the system ground state $|A_{0},0\rangle \approx |g,g_{a},0\rangle $ to
the state $|A_{1},1\rangle \approx -|g,e_{a},1\rangle $ as function of the
t-qubit's bare frequency $\Omega _{0}$ for the ancilla frequency $\Omega
_{a}=0.6\nu $. Figure \ref{fig1}b shows the resonant modulation frequency $%
\eta _{r}=E_{1,1}-E_{0,0}$ ($r$ and $\eta _{r}$ were obtained by
diagonalizing numerically $\hat{H}_{0}$). This transition can be roughly
interpreted as the Anti-Jaynes-Cummings ancilla--field regime in which one
photon and one ancilla excitation are created, while the t-qubit remains
approximately in the ground state. We verified that the relative errors
between the exact numeric results and the analytic expressions of Section %
\ref{ana} is below $0.6\%$ for the modulation frequency $\eta _{r}$ and
below $5\%$ for the transition rate $r$ (data not shown). In Figure \ref%
{fig1}c we show the fidelities $F_{k,n}(t)=|\langle A_{k},n|\psi (t)\rangle
|^{2}$ as function of time for ($k=0,n=0$) and ($k=1,n=1$), obtained by
solving numerically the SE for the original
Hamiltonian (\ref{tot}) with parameters $\Omega _{0}=0.95\nu $, $\eta
=1.586\nu $ and initial state $|g,g_{a},0\rangle $. For these parameters,
the weights in the state $|A_{1},1\rangle $ [Equation (\ref{weight}) with
properly chosen $i$] are $\phi _{2,i}^{(0)}=-0.168$, $\phi
_{2,i}^{(1)}=-0.977$, $\phi _{2,i}^{(2)}=0.021$ and $\phi _{2,i}^{(3)}=0.132$
(so the contribution of the bare state $|A_{1}^{1}\rangle $ is the largest)
and the transition rate is $r=1.93\times 10^{-4}$. As predicted in Section (%
\ref{ana}), there is a periodic population exchange between the
dressed-states $|A_{0},0\rangle $ and $|A_{1},1\rangle $ with period $\pi
/\nu r$, while other states remain practically unpopulated. In Figure \ref%
{fig1}d we present the numeric solution of the master equation (\ref{ssme1}%
). We plot the average photon number $\langle \hat{n}\rangle $, the qubits
excitations $\langle \hat{\sigma}_{e}\rangle $ and $\langle \hat{\sigma}%
_{e}^{(a)}\rangle $ and the average total number of excitations $%
N_{tot}=\langle (\hat{n}+\hat{\sigma}_{e}+\hat{\sigma}_{e}^{(a)})\rangle $.
For comparison, the average total number of excitation during unitary
evolution (labeled $N_{tot}^{free}$) is also shown. This figure attests that
photon generation from vacuum can occur even in the presence of dissipation.
The order of magnitude of maximal allowed dissipation rates (denoted as $%
\gamma _{\max }$) can be estimated from the condition $\gamma _{\max }/\nu
\sim r$, from which we obtain the value $\gamma _{\max }/g\sim $ $4\times
10^{-3}$ that is ubiquitous in several circuit QED setups \cite{63,64,65}.

\begin{figure}[tbh]
\begin{center}
\includegraphics[width=0.49\textwidth]{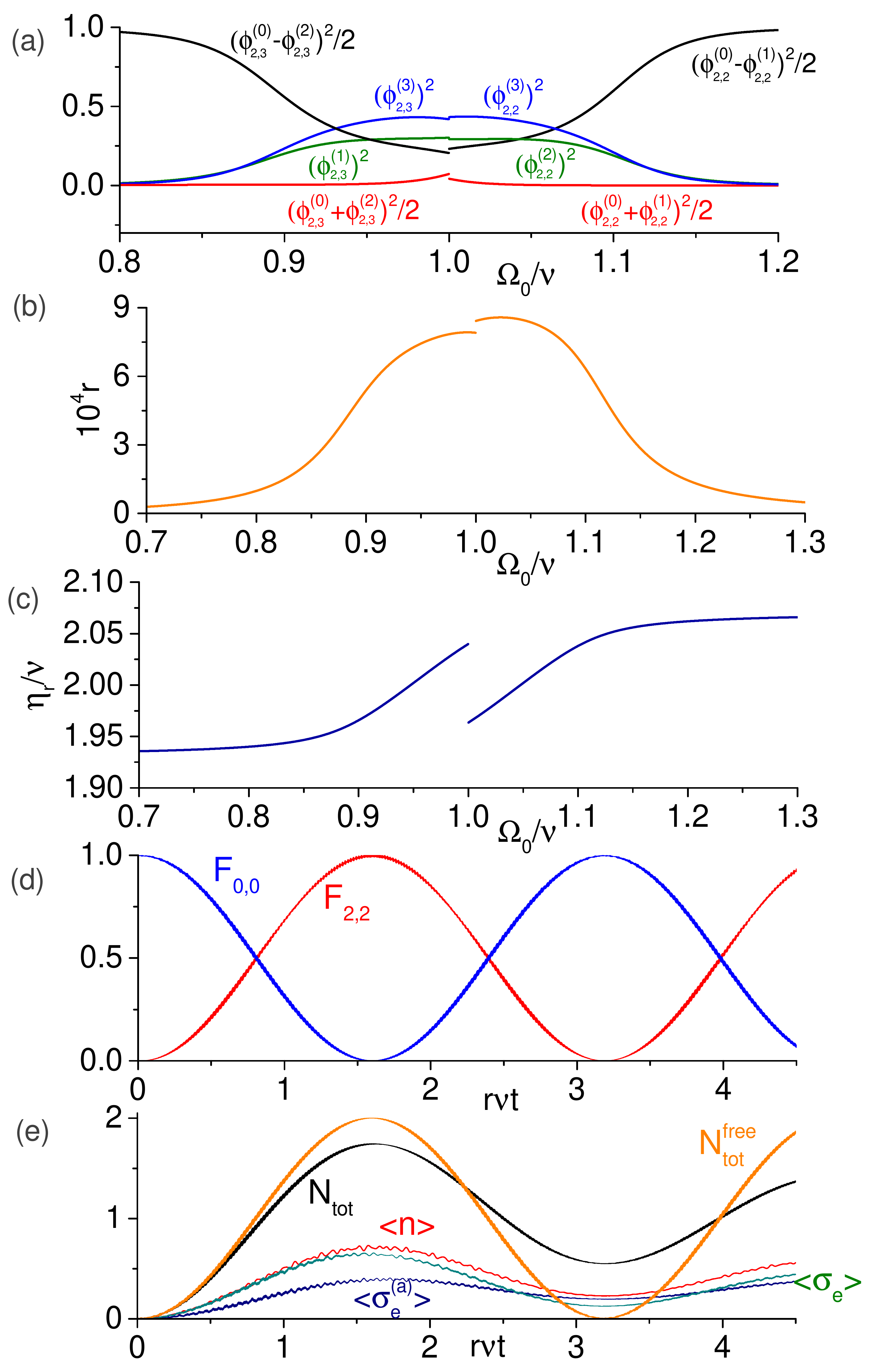} {}
\end{center}
\caption{Transition from $|A_{0},0\rangle $ to the 2-excitations states $|%
\protect\phi _{2,i}\rangle $ ($i=3$ for $\Omega _{0}<\protect\nu $, $i=2$
for $\Omega _{0}>\protect\nu $) for resonant ancilla with frequency $\Omega
_{a}=\protect\nu $. a) Composition of the dressed-states (\protect\ref%
{weight}) involved in the studied transition. b) Dimensionless transition
rate $r=|R_{0,0;\protect\phi _{2},i}|/\protect\nu $ as function of t-qubit's
bare frequency. c) Resonant modulation frequency. d) Fidelities of the
states $|A_{0},0\rangle $ and $|\protect\phi _{2,2}\rangle $ obtained from
numeric solution of the SE for $\Omega _{a}=\protect%
\nu $ and $\Omega _{0}=1.05\protect\nu $. e) Numeric dynamics of the average
numbers of excitations ($\langle n\rangle ,\langle \protect\sigma %
_{e}\rangle ,\langle \protect\sigma _{e}^{(a)}\rangle ,N_{tot}$) in the
presence of dissipation, compared to the average total number of excitations
$N_{tot}^{free}$ under unitary evolution. }
\label{fig2}
\end{figure}

\begin{table}[tbp]
\begin{tabular}{|l|l|l|l|l|}
\hline
t-qubit frequency & $\phi _{2,3}^{\left( 0\right) }$ & $\phi _{2,3}^{\left(
1\right) }$ & $\phi _{2,3}^{\left( 2\right) }$ & $\phi _{2,3}^{\left(
3\right) }$ \\ \hline
$\Omega _{0}/\nu =0.5$ & 0.735 & -0.017 & -0.678 & 0.012 \\ \hline
$\Omega _{0}/\nu =0.7$ & 0.741 & -0.048 & -0.669 & 0.038 \\ \hline
$\Omega _{0}/\nu =0.8$ & 0.740 & -0.122 & -0.653 & 0.105 \\ \hline
$\Omega _{0}/\nu =0.85$ & 0.714 & -0.240 & -0.618 & 0.227 \\ \hline
$\Omega _{0}/\nu =0.9$ & 0.579 & -0.449 & -0.479 & 0.484 \\ \hline
$\Omega _{0}/\nu =0.95$ & 0.457 & -0.538 & -0.312 & 0.636 \\ \hline
$\Omega _{0}/\nu =0.99$ & 0.489 & -0.548 & -0.178 & 0.655 \\ \hline
\end{tabular}%
\caption{Probability amplitudes in the state $|\protect\phi _{n=2,i=3}\rangle
$, Equation (\protect\ref{weight}), for $\Omega _{a}=\protect\nu $ and $\Omega
_{0}<\protect\nu $.}
\label{t1}
\end{table}

\begin{table}[tbp]
\begin{tabular}{|l|l|l|l|l|}
\hline
t-qubit frequency & $\phi _{2,2}^{\left( 0\right) }$ & $\phi _{2,2}^{\left(
1\right) }$ & $\phi _{2,2}^{\left( 2\right) }$ & $\phi _{2,2}^{\left(
3\right) }$ \\ \hline
$\Omega _{0}/\nu =1.01$ & -0.464 & 0.236 & 0.541 & 0.660 \\ \hline
$\Omega _{0}/\nu =1.05$ & -0.437 & 0.351 & 0.539 & 0.629 \\ \hline
$\Omega _{0}/\nu =1.1$ & -0.571 & 0.524 & 0.435 & 0.458 \\ \hline
$\Omega _{0}/\nu =1.15$ & -0.697 & 0.659 & 0.206 & 0.193 \\ \hline
$\Omega _{0}/\nu =1.2$ & -0.714 & 0.688 & 0.098 & 0.084 \\ \hline
$\Omega _{0}/\nu =1.3$ & -0.711 & 0.702 & 0.035 & 0.028 \\ \hline
$\Omega _{0}/\nu =1.5$ & -0.704 & 0.710 & 0.01 & 0.008 \\ \hline
\end{tabular}%
\caption{Probability amplitudes in the state $|\protect\phi _{n=2,i=2}\rangle
$ for $\Omega _{a}=\protect\nu $ and $\Omega _{0}>\protect\nu $.}
\label{t2}
\end{table}

In Figure \ref{fig2} we consider the ancilla at exact resonance with the
cavity mode, $\Omega _{a}=\nu $, and study the transition from $%
|A_{0},0\rangle $ to the state $|\phi _{n=2,i}\rangle $ given by Equation (\ref%
{weight}), in which $i=3$ for $\Omega _{0}<\nu $ and $i=2$ for $\Omega
_{0}>\nu $ (recall that the index $n$ specifies the subspace $\mathcal{A}_{n}
$). Tables \ref{t1} and \ref{t2} show the values of the probability amplitudes $\phi
_{2,i}^{\left( j\right) }$ ($j=0,\cdots ,3$) for several values of $\Omega
_{0}/\nu $ for parameters $\Omega _{a}=\nu $, $g=h=0.05\nu $. Figure \ref%
{fig2}a shows the quantities $(\phi _{2,3}^{\left( 0\right) }+\phi
_{2,3}^{\left( 2\right) })^{2}/2$, $(\phi _{2,3}^{\left( 0\right) }-\phi
_{2,3}^{\left( 2\right) })^{2}/2$, $(\phi _{2,3}^{(1)})^{2}$ and $(\phi
_{2,3}^{(3)})^{2}$ for $\Omega _{0}<\nu $ and $(\phi _{2,2}^{\left( 0\right)
}+\phi _{2,2}^{\left( 1\right) })^{2}/2$, $(\phi _{2,2}^{\left( 0\right)
}-\phi _{2,2}^{\left( 1\right) })^{2}/2$, $(\phi _{2,2}^{(2)})^{2}$ and $%
(\phi _{2,2}^{(3)})^{2}$ for $\Omega _{0}>\nu $. From Equation (\ref{no}) we see
that far from the resonance, $|\Omega _{0}-\nu |\gtrsim 0.2\nu $, the system
eigenstates are approximately $|\phi _{2,3}\rangle \approx |g\rangle \otimes
(|g_{a},2\rangle -|e_{a},1\rangle )/\sqrt{2}$ for $\Omega _{0}<\nu $ and $%
|\phi _{2,2}\rangle \approx -|g\rangle \otimes (|g_{a},2\rangle
+|e_{a},1\rangle )/\sqrt{2}$ for $\Omega _{0}>\nu $. Figures \ref{fig2}b and
\ref{fig2}c show the numeric results for the dimensionless transition rate $%
r=|R_{0,0;\phi_2,i}|/\nu $ ($i=3$ for $\Omega _{0}<\nu $ and $i=2$ for $%
\Omega _{0}>\nu $) and resonant modulation frequency $\eta
_{r}=E_{2,i}^{\phi }-E_{0,0}$ (where $E_{2,i}^{\phi }$ is the numeric
eigenvalue of $\hat{H}_{0}$ corresponding $\lambda _{2,i}^{\phi }$). The
agreement with analytic results of Section \ref{ana} is excellent, with the
relative error below 3\% for $r$ and below $0.1\%$ for $\eta _{r}$ (data not
shown). Figure \ref{fig2}d shows the numeric solution of the SE for the
fidelities $F_{0,0}(t)=|\langle A_{0},0|\psi (t)\rangle |^{2}$ and $%
F_{2,2}(t)=|\langle \phi _{2,2}|\psi (t)\rangle |^{2}$ for parameters $%
\Omega _{a}=\nu $, $\Omega _{0}=1.05\nu $ and $\eta =2.002\nu $, when the
transition rate is $r=8.4\times 10^{-4}$ and the weights in Equation (\ref{weight}%
) are $\phi _{2,2}^{(0)}=-0.437$, $\phi _{2,2}^{(1)}=0.351$, $\phi
_{2,2}^{(2)}=0.539$ and $\phi _{2,2}^{(3)}=0.629$ (see table \ref{t1}). As
expected, only the tripartite entangled state $|\phi _{2,2}\rangle $ becomes
periodically populated during the evolution. The panel \ref{fig2}e
illustrates the behavior of the average excitations numbers of the qubits
and the field in the presence of dissipation, together with the average
total number of excitations $N_{tot}$ and $N_{tot}^{free}$ (under unitary
evolution), confirming the feasibility of photon generation.

\begin{figure}[tbh]
\begin{center}
\includegraphics[width=0.49\textwidth]{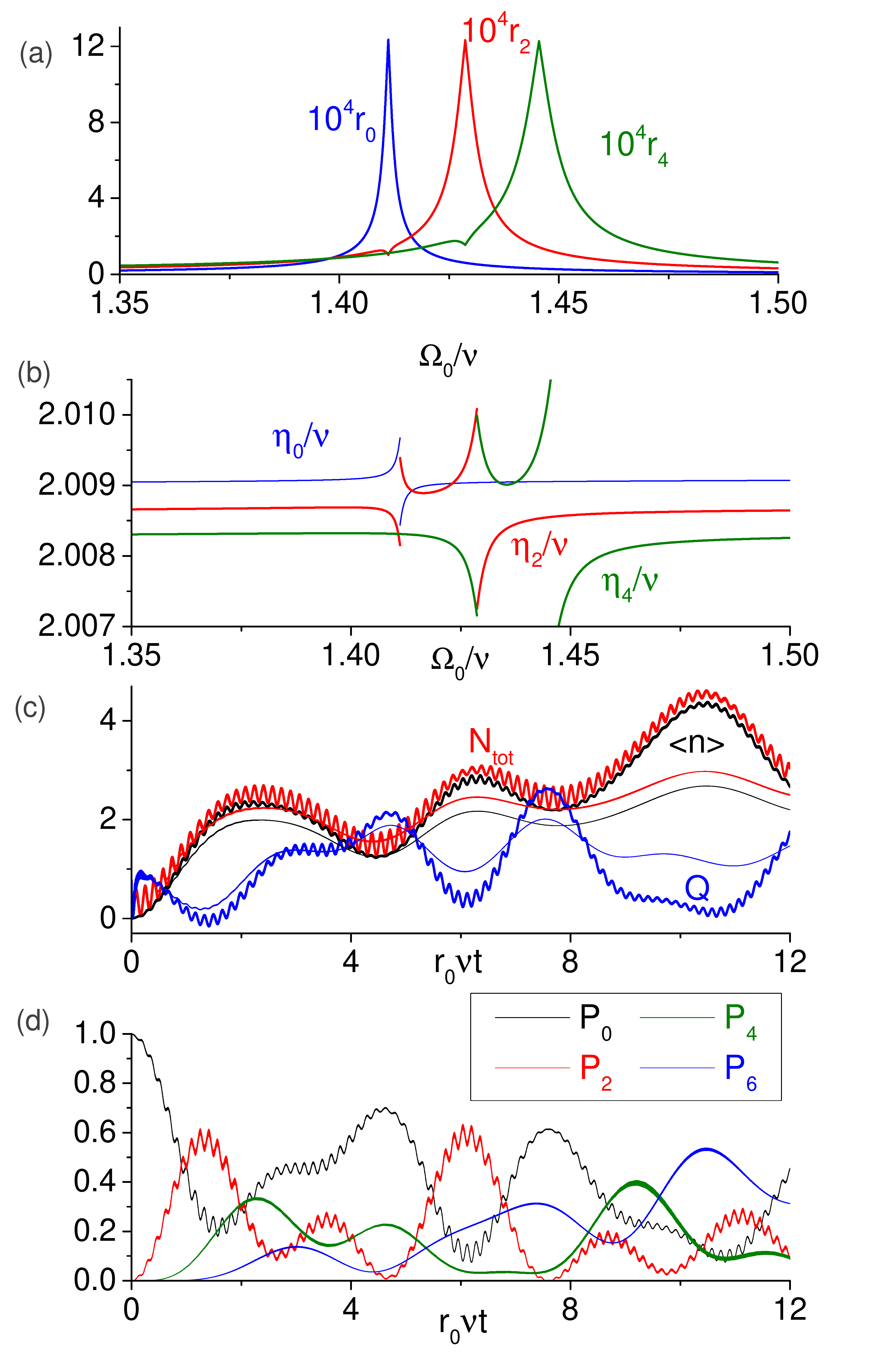} {}
\end{center}
\caption{DCE-like transition in which $|A_{0},n\rangle $ couples to the
state $|A_{0},n+2\rangle $ with even $n$ for off-resonant ancilla with $%
\Omega _{a}=0.6\protect\nu $. a) Dimensionless transition rates $%
r_{n}=|R_{0,n;0,n+2}|/\protect\nu $. b) Resonant modulation frequencies,
where discontinuities occur due to the modification of the dressed-states at
the degeneracy points. c) Numeric dynamics of the average photon number,
total number of excitations and the Mandel's Q-factor. Bold (thin) lines
denote the unitary (dissipative) evolution. Notice that even with
dissipation the generation of several excitations is possible. d) Dynamics
of the cavity Fock states with populations above 10\% during unitary
evolution. In this example, up to six photons can be generated with
substantial probabilities, while the probability of the vacuum state can
decrease below 10\%.}
\label{fig3}
\end{figure}

In Figure \ref{fig3} we analyze the possibility of generation of several
photons from vacuum due to intrinsic coupling between the states $%
|A_{0},n\rangle $ and $|A_{0},n+2\rangle $ near the degeneracy point $\Omega
_{0}+\Omega _{a}\approx 2\nu $, when both qubits are far detuned from the
cavity (in this regime $|A_{0},n\rangle \approx |g,g_{a},n\rangle $ and $%
n=0,2,4,\cdots ,n_{\max }$, where $n_{\max }$ denotes the most excited Fock
state for a given modulation frequency). Figure \ref{fig3}a shows the
dimensionless transition rates $r_{n}=|R_{0,n;0,n+2}|/\nu $ for $n=0,2,4$ as
function of $\Omega _{0}/\nu $, while Fig \ref{fig3}b show the resonant
modulation frequencies $\eta _{n}=E_{0,n+2}-E_{0,n}$ for the same parameters
as in Figure \ref{fig1}. The relative error of our analytic formulae is
below 5\% (0.1\%) for the transition rates (modulation frequencies). A
single discontinuity in $\eta _{0}$ and two discontinuities in $\eta _{2}$
and $\eta _{4}$ are expected, since the eigenvalues $E_{0,n\geq 2}$ present
one discontinuity near the degeneracy point (while $E_{0,0}$ is continuous).
From these figures we infer that at least three states $|A_{0},n\rangle $
with $n=2,4,6$ could be populated from the initial ground state provided the
modulation frequency is sufficiently close to all the three frequencies $%
\eta _{0},\eta _{2}$ and $\eta _{4}$ (with the mismatch smaller or of the
order of $\nu r_{n}$). This hint is confirmed in Figure \ref{fig3}c, where
we solved numerically the SE and the master equation for parameters $\Omega
_{a}=0.6\nu $, $\Omega _{0}=1.405\nu $ and $\eta =2.0086\nu $, when the
transition rates are approximately $r_{0}\approx 1.8\times 10^{-4}$, $%
r_{2}\approx 1.1\times 10^{-4}$ and $r_{4}\approx 9.6\times 10^{-5}$. This
plot displays the average photon number, the average total excitation number
and the Mandel's Q-factor $Q=[\langle (\Delta \hat{n})^{2}\rangle -\langle
\hat{n}\rangle ^{2}]/\langle \hat{n}\rangle $ (that quantifies the spread of
the photon number distribution); bold (thin) lines depict the unitary
(dissipative) evolution. We see that a small number of photons can be
created from vacuum even in the presence of dissipation, and the qubits
remain approximately in the ground states because $N_{tot}$ is always close
to $\langle \hat{n}\rangle $. The behavior of $\langle \hat{n}\rangle $ is
better understood by looking at Figure \ref{fig3}d, in which we plot the
photon number probabilities of the Fock states, $P_{n}=\mathrm{Tr}[\hat{\rho}|n\rangle \langle
n|]$, with occupation probabilities above 10\% under the
unitary evolution. We see that up to six photons can be generated with
significant probabilities, and the populations of the Fock states exhibit
irregular oscillations due to the simultaneous coupling between the states $%
|A_{0},2n\rangle $ with $n=0,\ldots ,3$. We also observe that the created
field state is very different from the squeezed vacuum state generated in standard
single-mode cavity DCE with vibrating walls or time-dependent permittivity
\cite{rev1}, for which $Q=1+2\langle \hat{n}\rangle $.

\subsection{Multi-photon transitions}

\label{fin}

For larger values of the light-matter coupling constant $g$ (ultrastrong
coupling regime \cite{m1,m2}), generation of photons from vacuum becomes
feasible via direct driving of the ground state of the Hamiltonian $\hat{H}_{0}
$ to excited eigenstates with $n>2$ excitations. Here we illustrate this rich
variety of phenomena by considering the transitions from $|A_{0},0\rangle $
to the states $|A_{0},4\rangle $ and $|A_{0},6\rangle $. For the reasons
already mentioned at the end of Subsection \ref{fop}, the transition rates and resonant
modulation frequencies were obtained by diagonalizing numerically $\hat{H}%
_{0}$.

\begin{figure}[tbh]
\begin{center}
\includegraphics[width=0.49\textwidth]{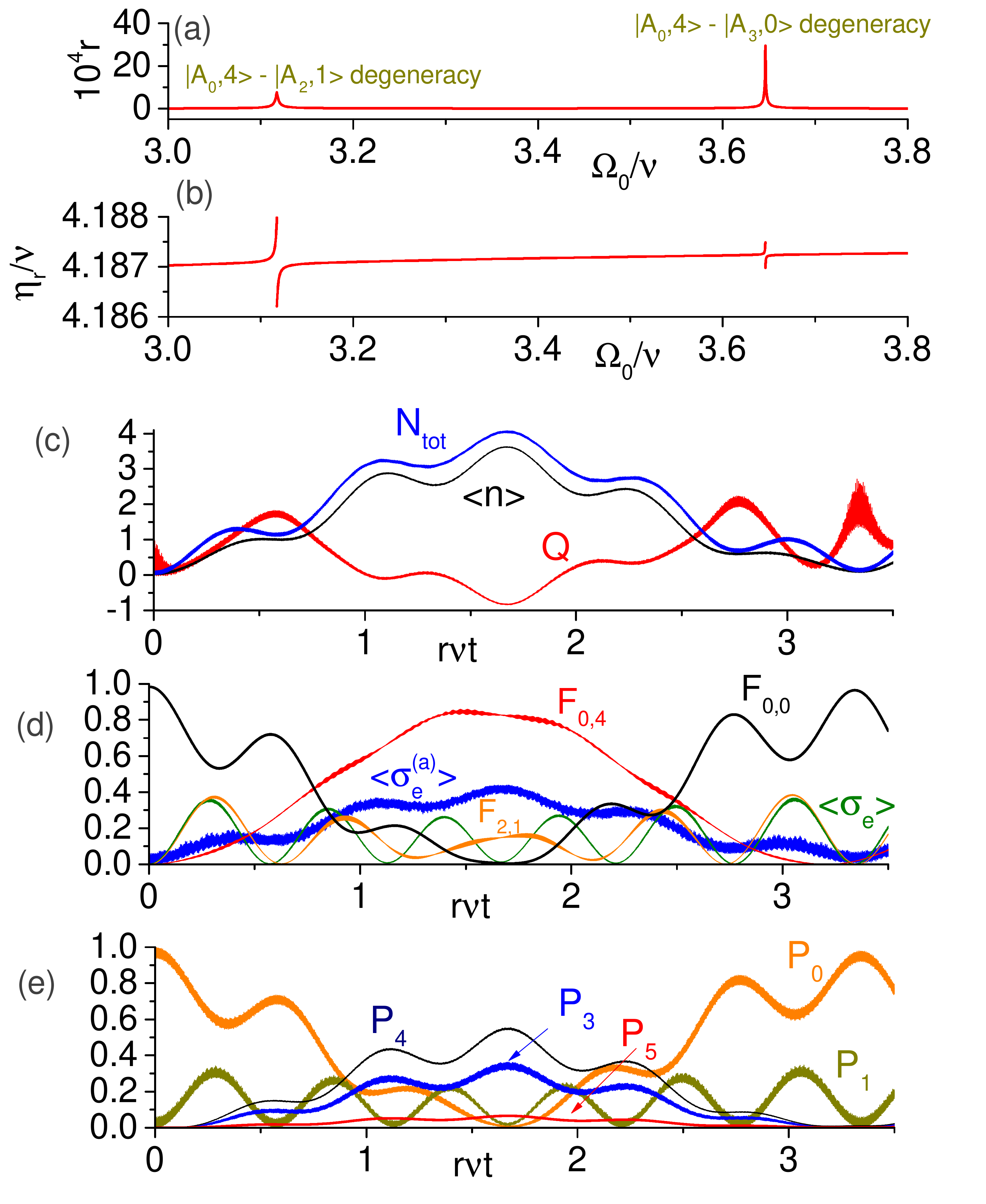} {}
\end{center}
\caption{Numeric results for the direct transition $|A_{0},0\rangle
\rightarrow |A_{0},4\rangle $ in the ultrastrong coupling regime for
parameters $\Omega _{a}=0.6\protect\nu $, $g=0.2\protect\nu $, $h=0.1\protect%
\nu $. a) Dimensionless transition rate $r=|R_{0,0;0,4}|/\protect\nu $; the
peaks occur near degeneracies between the states $\{|A_{0},4\rangle
,|A_{2},1\rangle \}$ and $\{|A_{0},4\rangle ,|A_{3},0\rangle \}$. b)
Resonant modulation frequency $\protect\eta _{r}=E_{0,4}-E_{0,0}$. c)
Average photon number, average total number of excitations and the Mandel's
Q-factor for $\Omega _{0}=3.12\protect\nu $ and $\protect\eta =4.1873\protect%
\nu $ under unitary evolution. d) Populations of t-qubit and ancilla;
fidelities of the states $|A_{0},0\rangle $, $|A_{0},4\rangle $ and $%
|A_{2},1\rangle $. e) Dynamics of the Fock states with occupation
probabilities above 5\%.}
\label{fig4}
\end{figure}

In Figure \ref{fig4}a we set the parameters $\Omega _{a}=0.6\nu $, $g=0.2\nu
$ and $h=0.1\nu $, and evaluate numerically the dimensionless transition rate $%
r=|R_{0,0;0,4}|/\nu $ for the transition $|A_{0},0\rangle \rightarrow
|A_{0},4\rangle $. As in previous figures, the transition rate presents
sharp peaks near the degeneracies between $|A_{0},4\rangle $ and the states $%
|A_{2},1\rangle $ or $|A_{3},0\rangle $. The corresponding resonant
modulation frequency $\eta _{r}=E_{0,4}-E_{0,0}$ is shown in Figure \ref%
{fig4}b. To confirm the feasibility of such multi-photon transition, we
solved numerically the SE for $\Omega _{0}=3.12\nu $ and $\eta =4.1873\nu $,
when the transition rate is $r\approx 3.2\times 10^{-4}$, the ground state
is $|A_{0},0\rangle \approx 0.99|A_{0}^{0}\rangle +0.13|A_{1}^{1}\rangle
+0.02|A_{0}^{2}\rangle $ and the near degenerate dressed-states are $|0,4\rangle
\approx 0.72|A_{0}^{4}\rangle -0.57|A_{1}^{3}\rangle -0.27|A_{2}^{1}\rangle
+0.25|A_{1}^{5}\rangle +0.12|A_{3}^{0}\rangle $ (with approximate energy $%
4.1868\nu $ above the ground state energy) and $|2,1\rangle \approx
0.86|A_{2}^{1}\rangle -0.39|A_{3}^{0}\rangle +0.22|A_{0}^{4}\rangle
-0.18|A_{1}^{3}\rangle +0.16|A_{3}^{2}\rangle $ (with the corresponding
energy $4.1899\nu $). Figure \ref{fig4}c shows the behavior of $\langle \hat{%
n}\rangle $, the average total number of excitations and the Mandel's
factor, while Figure \ref{fig4}d shows the excitation probability of t-qubit
and ancilla for the initial zero-excitation state $|g,g_{a},0\rangle $.
This complicated behavior arises because the modulation couples the ground
state $|A_{0},0\rangle $ to both the states $|A_{0},4\rangle $ and $%
|A_{2},1\rangle $, as attested by the fidelities $F_{k,l}=|\langle A_{k},l |\psi
(t)\rangle|^{2}$ shown in \ref{fig4}d (we have $%
F_{0,0}+F_{0,4}+F_{2,1}>0.98$, proving that only the states $|A_{0},0\rangle
$, $|A_{0},4\rangle $ and $|A_{2},1\rangle $ are significantly populated).
We see that nearly 4 excitations are created during the expected time
interval $r\nu t=\pi /2$, and both qubits can be found in excited states
with significant probabilities. The excitation of the ancilla comes mainly
from the contribution of the state $|A_{1}^{3}\rangle $, and the excitation
of the t-qubit comes mainly from the contribution of $|A_{2}^{1}\rangle $ in
the dressed-states (since $|A_{1}\rangle \approx -|g,e_{a}\rangle $ and $%
|A_{2}\rangle \approx |e,g_{a}\rangle $ for the chosen parameters). Up to
five photons can be detected with probabilities above 5\%, and the
probability of the zero-photon state becomes nearly zero for $r\nu t\approx
\pi /2$, as illustrated in Figure \ref{fig4}e.

\begin{figure}[tbh]
\begin{center}
\includegraphics[width=0.49\textwidth]{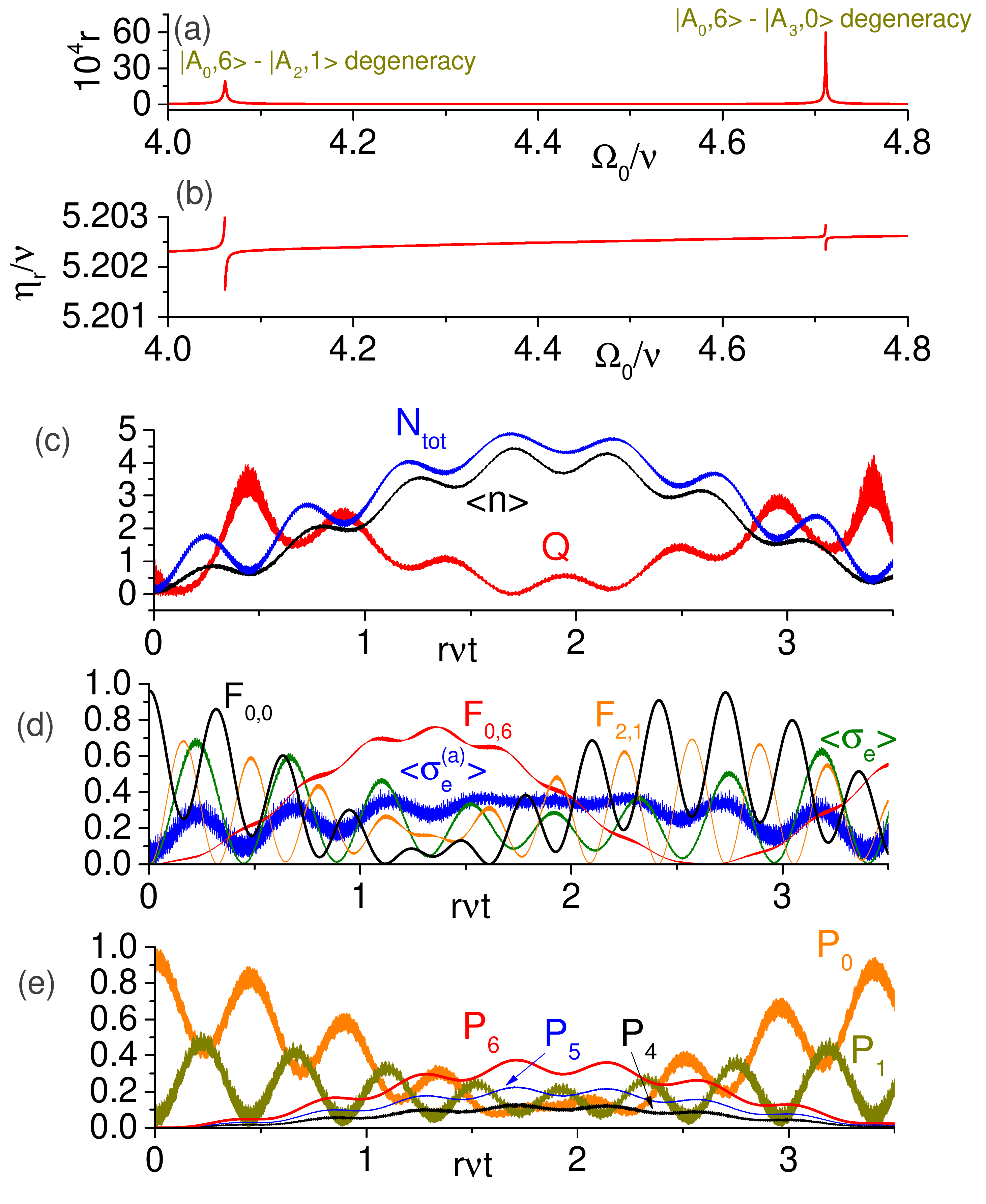} {}
\end{center}
\caption{Numeric results for the direct transition $|A_{0},0\rangle
\rightarrow |A_{0},6\rangle $ in the ultrastrong coupling regime for
parameters $\Omega _{a}=0.6\protect\nu $, $g=0.3\protect\nu $, $h=0.2\protect%
\nu $. a) Dimensionless transition rate $r=|R_{0,0;0,6}|/\protect\nu $; the
peaks occur near degeneracies between the states $\{|A_{0},6\rangle
,|A_{2},1\rangle \}$ and $\{|A_{0},6\rangle ,|A_{3},0\rangle \}$. b)
Resonant modulation frequency $\protect\eta _{r}=E_{0,6}-E_{0,0}$. c)
Average photon number, average total number of excitations and the Mandel's
Q-factor for $\Omega _{0}=4.057\protect\nu $ and $\protect\eta =5.201\protect%
\nu $ under unitary evolution. d) Populations of t-qubit and ancilla;
fidelities of the states $|A_{0},6\rangle $, $|A_{2},1\rangle $ and $%
|A_{0},0\rangle $. e) Dynamics of the Fock states with occupation
probabilities above 10\%.}
\label{fig5}
\end{figure}

Figure \ref{fig5} is analogous to Figure \ref{fig4} but for the direct
transition $|A_{0},0\rangle \rightarrow |A_{0},6\rangle $. We performed
numerical simulations for larger coupling strengths $g=0.3\nu $ and $%
h=0.2\nu $, but the same ancilla frequency $\Omega _{a}=0.6\nu $ as in
Figures \ref{fig3} and \ref{fig4}. In panel \ref{fig5}a we verify a strong
enhancement of the transition rate $r=|R_{0,0;0,6}|/\nu $ in the vicinity of
degeneracy between $|A_{0},6\rangle $ and the states $|A_{2},1\rangle $ or $%
|A_{3},0\rangle $. The resonant modulation frequency $\eta
_{r}=E_{0,6}-E_{0,0}$ is shown in Figure \ref{fig5}b. In panels \ref{fig5}%
c-e we solved numerically the SE for the initial state $|g,g_{a},0\rangle $ and
parameters $\Omega _{0}=4.057\nu $ and $\eta =5.201\nu $, when the near
degenerate states are $|A_{0},6\rangle \approx 0.68|A_{0}^{6}\rangle
+0.52|A_{1}^{5}\rangle -0.39|A_{0}^{4}\rangle +0.25|A_{1}^{7}\rangle
+0.16|A_{1}^{3}\rangle -0.14|A_{2}^{1}\rangle $ (with approximate energy $%
5.2025\nu $ above the ground state energy) and $|A_{2},1\rangle \approx
0.8|A_{2}^{1}\rangle -0.52|A_{3}^{0}\rangle +0.25|A_{3}^{2}\rangle
+0.11|A_{0}^{6}\rangle $ (with energy $5.1978\nu $), while the ground state
is $|A_{0},0\rangle \approx 0.98|A_{0}^{0}\rangle +0.19|A_{1}^{1}\rangle
+0.04|A_{0}^{2}\rangle $. Figure \ref{fig5}c confirms the generation of
photons with $\langle \hat{n}\rangle \leq 4.5$ and $N_{tot}\leq 5$; the
populations of Fock states with occupation probabilities above 10\% are
shown in Figure \ref{fig5}e. As expected, up to 6 photons are generated with
significant probabilities (while the Fock states $|2\rangle $ and $|3\rangle
$ are practically unpopulated), and the substantial population of 1-photon
state comes from the partial excitation of the dressed-state $%
|A_{2},1\rangle $. Figure \ref{fig5}d shows the fidelities of the dressed
states $|A_{0},6\rangle $, $|A_{2},1\rangle $ and $|A_{0},0\rangle $, whose
sum is always above 96\%, attesting that only these states become
significantly populated during the evolution.

\begin{figure}[tbh]
\begin{center}
\includegraphics[width=0.49\textwidth]{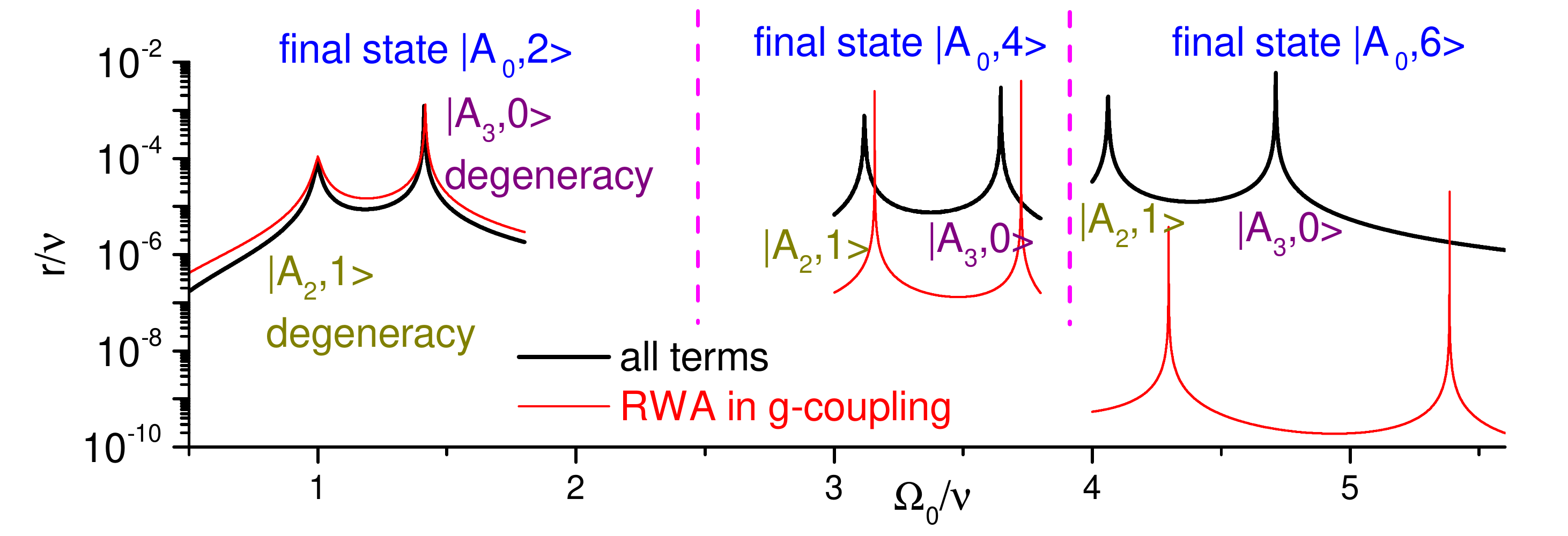} {}
\end{center}
\caption{Comparison of the direct transition rates from the initial state $%
|A_{0},0\rangle $ to the states $|A_{0},2\rangle $, $|A_{0},4\rangle $ and $%
|A_{0},6\rangle $ (indicated on top) with the complete Hamiltonian $\hat{H}%
_{0}$ (black lines) and under RWA with $g( \hat{a}\hat{\protect\sigma}%
_{-}^{(a)}+h.c.) =0$ (red lines). The parameters are the same as in
Figures \protect\ref{fig3}, \protect\ref{fig4} and \protect\ref{fig5},
respectively. The peaks occur near degeneracies between the states \{$%
|A_{0},n\rangle ,|A_{2},1\rangle $\} and \{$|A_{0},n\rangle ,|A_{3},0\rangle
$\} for $n=2,4,6.$}
\label{fig6}
\end{figure}

To conclude, we emphasize that in the regimes studied above the
ancilla-cavity counter-rotating term $g(\hat{a}\hat{\sigma}_{-}^{(a)}+h.c.)$
in the Hamiltonian $\hat{H}$ plays an essential role. In the 2-excitations
transitions studied in figures \ref{fig1} - \ref{fig3} its major effect is
to alter the resonant modulation frequencies and the position of peaks
in transition rates, so in the first approximation it could be neglected via
the Rotating Wave Approximation (RWA) (the transition rate would be wrong by
roughly 30\% in this case, provided one adjusted by hand the positions of
the peaks). However, for 4- and 6-excitations transitions studied in Figures %
\ref{fig4} and \ref{fig5} it is indispensable, and the transition rates
would be orders of magnitude smaller if it were neglected. This is
illustrated in Figure \ref{fig6}, where we plot the transition rates of
Figures \ref{fig3}a, \ref{fig4}a and \ref{fig5}a (corresponding to the
direct transition $|A_{0},0\rangle \rightarrow |A_{0},n\rangle $, $n=2,4,6$)
with and without the counter-rotating term and indicate the origin of each
peak (degeneracy of $|A_{0},n\rangle $ with $|A_{2},1\rangle $ or $%
|A_{3},0\rangle $).

\section{Conclusions\label{conc}}

We showed that photons can be generated from vacuum inside a stationary
resonator due to resonant time-modulation of a quantum system
\textquotedblleft invisible\textquotedblright\ to the Electromagnetic field,
i. e., when it is indirectly coupled to the field via some ancilla quantum
subsystem. Such coupling may be advantageous when the modulated system is
located outside or at the end of the resonator (when its coupling to the
field is zero) to reduce the influence of the external driving on the cavity
field, while it is strongly coupled to one or several ancilla subsystems.
Moreover, this setup permits fabrication of quantum systems designed
specifically for rapid external modulation at the cost of increased
dissipation, while the stationary ancillas and cavities can be optimized for
minimal dissipative losses.

We considered the simplest scenario of a single-mode cavity and a
time-modulated qubit (t-qubit) playing the role of the \textquotedblleft
invisible\textquotedblright\ system, while the ancilla consisted of a
stationary qubit dipole-coupled to both the cavity and the t-qubit. Our
scheme and mathematical analysis can be readily generalized to more complex
systems with alternative coupling mechanisms. We deduced analytically the
system dynamics under unitary evolution and showed that the rate of photon
generation can be drastically enhanced near certain resonances of the
tripartite system (related to anti-crossings in the energy spectrum). We
exemplified our scheme by studying photon generation from the initial vacuum
state of the system, considering regimes in which either a single
dressed-state or a small number of dressed-states with a few photons are
populated under sinusoidal modulation. Moreover, we demonstrated that these
phenomena persist in the presence of weak dissipation, with typical
transition rates of the order of $10^{-4}\nu $ (where $\nu $ is the cavity
frequency). Although the present study focused on creation of photons
from vacuum, it could find applications in the engineering of effective
interactions and generation of useful multipartite entangled states.

\section*{Acknowledgements}

A.V.D. acknowledges partial support from National Council for Scientific and
Technological Development -- CNPq (Brazil). W.W.T.S. and M.V.S.P.
acknowledge the support of ProIC/DPG/UnB and Coordena\c{c}\~{a}o de Aperfei%
\c{c}oamento de Pessoal de N\'{\i}vel Superior - Brasil (CAPES) -- Finance
Code 001, via program PET F\'{\i}sica UnB.

\end{document}